\documentstyle[preprint,aps,pre,epsf]{revtex}
\begin{document}
\def\bea{\begin{eqnarray}}
\def\eea{\end{eqnarray}}
\def\a{\alpha}
\def\b{\beta}
\def\d{\delta}
\def\l{\lambda}
\def\p{\partial} 
\def\#{\nonumber}
\def\r{\rho}
\def\la{\langle}
\def\ra{\rangle}
\def\e{\epsilon}
\def\n{\eta}
\def\hn{\hat{\eta}}
\def\break#1{\pagebreak \vspace*{#1}}
\def\byn{\frac{1}{n}}
\draft
%ver_jan30.tex
\title{ Solution of A Generalized Stieltjes Problem }
\author{B. Sriram Shastry $^1$ and Abhishek Dhar $^2$ }
\address{  $1$ Bell Laboratories Lucent
Technologies, 600 Murray Hill, NJ 07974 and \\
Indian Institute of Science, Bangalore 560012, India. \\
$^2$Raman Research Institute, Bangalore 560080, India and \\ Poornaprajna
Institute, Bangalore, India.\\}
\date{\today}
\maketitle
\widetext
\begin{abstract}
We present the exact solution for a set of nonlinear algebraic
equations $\frac{1}{z_l}= \pi d + \frac{2 d}{n} \sum_{m \neq l}
\frac{1}{z_l-z_m}$. These were
 encountered by us  in a recent study of the low energy
spectrum of the Heisenberg ferromagnetic chain \cite{dhar}. These  
equations are low $d$ (density) ``degenerations'' of more complicated
 transcendental equation of Bethe's Ansatz for a ferromagnet, but are
 interesting in themselves. They generalize, through a
single parameter, the  equations of Stieltjes, 
 $x_l = \sum_{m \neq l} 1/( x_l-x_m)$,  familiar from Random Matrix theory.
 It is shown that the solutions of these set of equations is 
given by the zeros of
generalized associated Laguerre polynomials. These zeros are interesting,
since they provide one of the few known cases where the location is along a 
nontrivial curve   in the complex plane that is determined in this work.
 Using a ``Green's function''
and a saddle point technique we determine the asymptotic distribution of zeros.
\end{abstract}

%\pacs{PACS numbers: 05.50.+q, 02.50.Ey, 05.40.-a}

%\narrowtext
%\textwidth=12cm

\section{Introduction}

The study of nonlinear algebraic sets of equations that arise in
 various physical contexts is a rich field of study. A famous example is 
the problem of Stieltjes, namely  the set of equations $x_l = \sum_{m
 \neq l} 1/( x_l-x_m)$, for $n$  
variables $x_l$, that arises when we consider the stationary points of the 
Random Matrix Gaussian Ensemble ``action''  $1/2 \sum x_l^2 - 1/2 \sum_{l
 \neq m} \log(|x_l-x_m|)$ \cite{mlmehta}. In that context, the well
 known result is that the $x_l$ are all real,
and further are the roots of Hermite's polynomials
 of degree  $n$,  forming a  dense set along the real line with the familiar
 semicircular density of states as $n \rightarrow \infty$. 

In a recent study of the ``most elementary excitations'' of the Heisenberg
 ferromagnet in 1-d \cite{dhar}, we came across a
one parameter  generalization of the Stieljes problem
\bea
\frac{1}{z_l}= \pi d + \frac{2 d}{n} \sum_{m \neq l}
\frac{1}{z_l-z_m}.
\label{ldeq}
\eea    
where the single parameter ``d'' ( $0\leq d\leq 1$) has the physical
 significance of the density of (hardcore) 
particles on a lattice. By a simple
 redefinition  of variables (see below), this is recognizable as a 
generalized Stieltjes problem $\frac{\b_l}{1+i \l \b_l}=\frac{1}{n} 
\sum_{m \ne l} \frac{1}{\b_l-\b_m}$, where, as the parameter $\l$ vanishes,
 it reduces to the standard problem. 

We found the equation quite fascinating in its own terms, requiring
some new tricks to solve. It further appears that
this problem connects with \cite{szego,lang,chenismail,askey,dette} 
that of  the asymptotic distribution 
of zeros of the standard  polynomials (orthogonal in most instances
though not in the present case). 
It  provides an  explicit example of a case where the zeros 
live on a nontrivial curve in the complex plane, which is determined here. 

The paper is organized as follows. In Sec. II we obtain a
perturbative solution of Eq.~(\ref{ldeq}) while in Sec. III we
give the exact solution. The construction of the ``Green's function'' and
the determination of distribution of zeros are described in 
Sec. IV. We summarize our results in Sec. V. Some results
of Stieltjes on the properties of zeros of orthogonal polynomials are
described in Appendix (A). In Appendix (B) we describe a second method
of obtaining the Green's function.

\section{The perturbative solution}

We  first present a ``perturbative solution'' in a parameter displayed below,
perturbing around the Stieltjes example.
In Eq.~(\ref{ldeq}) we make a change of variables $z_l=(1+i \lambda
\b_l)/{(\pi d)}$, with $\lambda=\sqrt{2 d}$. 
The new variable $\b_l$ satisfies the following equation:
\bea
\frac{\b_l}{1+i \l \b_l}=\frac{1}{n} \sum_{m \ne l} \frac{1}{\b_l-\b_m} 
\label{bldeq}
\eea
We note that for $\l=0$, this equation reduces to the well known form whose
solution can be found in terms of the roots of the Hermite polynomials,
 the example of Stieltjes \cite{stiel}. The general formalism for this
is summarized and expanded upon  in App.~(\ref{appa}). 
Thus one obtains $\b_l=x_l$ where $x_l$ satisfy $H_n(\sqrt{n} x_l)=0$,
where $H_n$ is the Hermite polynomial of degree $n$. The division by $n$ in 
Eq(\ref{bldeq})  is our preference, since it enables an easier passage to
the ``thermodynamic limit'''' $n\rightarrow \infty$, and we discuss this 
feature later.

One possible approach is to try for a perturbative solution around the
known result of Stieltjes to various orders in the small parameter
$\l$  (see below).
The perturbation analysis requires a knowledge of the eigenvectors 
and eigenvalues of an interesting matrix  studied  by Calogero (we
call this the Calogero matrix) in which the matrix elements are
algebraic functions of $x_l$, the roots of the Hermite polynomials.

For $\l \neq 0$ we try a perturbation series solution of the
form $\b_l =\sum_k \b_l^{(k)} (i \l)^k$. 
Putting this in Eq.~(\ref{ldeq}) and matching terms order by order we
get the following set of equations for the unknown coefficients
$\b_l^{(k)}$:
\bea
&\b_l^{(0)}&-\frac{1}{n} \sum_{m \neq l}
\frac{1}{\b_l^{(0)}-\b_m^{(0)}}=0 \# \\  
&\b_l^{(1)}& +\frac{1}{n} \sum_{m \neq l}
\frac{\b_l^{(1)}-\b_m^{(1)}}{[\b_l^{(0)}-\b_m^{(0)}]^2}=[\b_l^{(0)}]^2 \# \\
&\b_l^{(2)}&+\frac{1}{n} \sum_{m \neq l}
\frac{\b_l^{(2)}-\b_m^{(2)}}{[\b_l^{(0)}-\b_m^{(0)}]^2}=2 \b_l^{(1)}
\b_l^{(0)} -[\b_l^{(0)}]^3+\frac{1}{n} \sum_{m \neq l} \frac{[\b_l^{(1)}-\b_m^{(1)}]^2} 
{[\b_l^{(0)}-\b_m^{(0)}]^3} \# \\
&\b_l^{(3)}&+\frac{1}{n} \sum_{m \neq l}
\frac{\b_l^{(3)}-\b_m^{(3)}}{[\b_l^{(0)}-\b_m^{(0)}]^2}=2 \b_l^{(2)}
\b_l^{(0)} +[\b_l^{(1)}]^2-3 \b_l^{(1)} [\b_l^{(0)}]^2+[\b_l^{(0)}]^4 \# \\
&&~~~~~~~~~~~~~~~~~~~~~~~+\frac{2}{n} \sum_{m \neq l} \frac{[\b_l^{(2)}-\b_m^{(2)}][ \b_l^{(1)}-\b_m^{(1)}]}
{[\b_l^{(0)}-\b_m^{(0)}]^3}
-\frac{1}{n} \sum_{m \neq l} \frac{[\b_l^{(1)}-\b_m^{(1)}]^3} 
{[\b_l^{(0)}-\b_m^{(0)}]^4} \# \\
&&...
\label{pert}  
\eea
As noted above, the lowest order solution is given by $\b_l^{(0)}=x_l$
where $H_n(\sqrt{n} x_l)=0$. It turns out that it 
is possible to solve the equations of the perturbative series {\it at
every order}. To see this we first define the Calogero matrix \cite{calo}:
\bea 
T_{lm}=\d_{lm} \sum_{j \neq l}
\frac{1}{(x_l-x_j)^2}-(1-\d_{lm})\frac{1}{(x_l-x_m)^2} .
\eea
We then note that the set of equations in Eq.~(\ref{pert}), for $k >
0$, have the following general structure
\bea
\b_l^{(k)}+\frac{1}{n}\sum_m T_{lm} \b_m^{(k)}=g_l^{(k)}
\label{simpeq}
\eea
where the function $g_l^{(k)}$ appearing at $k$th order in
perturbation is a function of lower order terms only and so is
known. Hence at every stage we have to essentially solve a linear
matrix equation with the same matrix appearing at all orders.
This can be done by using some rather special properties of 
the  Calogero matrix, $T$. The following result is easily obtained.
\bea
T_{lm} x_m^r=rx_l^r-\frac{(r-1)(2 n-r)}{2}
x_l^{r-2}-\sum_{s=1}^{[r/2-1]} (r-2 s-1) (\sum_{m=1}^n x_m^{2 s})
x_l^{r-2 s-2}
\label{matp}
\eea  
where $[p]$ denotes the largest integer $ \le p$. Note that the rhs 
of Eq.~(\ref{matp}) involves powers of $x_l$ alone. It is also possible to
show that at every order the function $g_l^{(k)}$ is a known
polynomial of degree $(k+1)$ in the variable $x_l$. 
Thus it follows that the $k$th order solution can be obtained in the
form
\bea
\b_l^{(k)}= \sum_{r=0}^{k+1} c_r^{(k)} x_l^r,
\eea
where the coefficients $c_r^{(k)}$ are obtained by inserting the above form
into Eq.~(\ref{simpeq}) , using the property Eq.~(\ref{matp}) of the
Calogero matrix $T$ and  then comparing lhs and rhs of the equation. 
Till second order we get:
\bea
\b_l=x_l+i \l [ \frac{1}{3} x_l^2+ \frac{1}{3}(1-\frac{1}{n})]+ 
(i \l)^2 [\frac{1}{36} x_l^3+\frac{1}{72}(14-\frac{11}{n})
x_l] +O(\l^3) 
\label{ldroots}
\eea  
It is straightforward, though tedious, to carry the perturbation
to any order. We have been unable to find a closed form for the
general term.

As an example of this perturbation theory we note that the
 first order terms in $\l$ 
is already enough to compute several objects of interest. In the 
ferromagnet problem, one needs the ``energy'' of the state $w$ , defined by
$ w=4/(n d) \sum_l 1/ z_l^2$.( The present
 $w= n \epsilon/dJ$ or Ref(\cite{dhar})).
To obtain this energy to order $d^2$ the perturbation series till first
order in $\l$ suffices. 
In the limit of large $n$ the $x_l$ form a continuum stretching from
$- \sqrt{2}$ to $\sqrt{2}$ with the familiar semicircular 
density of states $\rho(x)= \frac{1}{\pi} \sqrt{2 - x^2}$.
 This solution can be used
to obtain the energy to order $d^2$. 
The energy $w $  for low $d$ can be
 found as $w = 4  \pi^2 [ d 
- d^2 \{ 6 <(\beta_l^{(0)} )^2> +
4 < \beta_l^{(1)} > \} + O( d^3) ]$, where the averages are normalized
sums over the indicated variables.
 Using the explicit expression Eq~(\ref{ldroots}) and  converting 
the sums to integrals
 over the  semicircular density of states we get finally the low
density formula 	
\bea
w= 4 \pi^2 d (1-d) + O(d^3).
\eea

\section{Exact solution}
The exact solution of Eq.~(\ref{ldeq}) can be obtained
using the results of Stieltjes which are described in App.~(\ref{appa}).  
We will give two different but related solutions.
The first one is directly related to the perturbative approach. We first
note that after change of variables $u_l=\sqrt{n} \b_l$, Eq.~(\ref{bldeq})
becomes of the form Eq.~(\ref{genst}) with $p(u)=(1+i {(2d/n)}^{1/2}
u);~q(u)=-2u$. The $\{u_l\}$ are then obtained as zeros of the
the corresponding polynomial function $g(u)$ which in this case
satisfies the following differential equation: 
\bea
(1+i \nu u) g''(u) -2 u g(u) +2 n g(u) = 0, 
\eea
  where $\nu= \sqrt{2d/n}$. For $\nu=0$ this is just Hermite's
equation and we recover our zeroth order perturbative result. For $\nu
\neq 0 $, we obtain the following series  solutions (which we denote 
by $Q_n(u)$):
\bea
Q_n(u) =\sum_{k=0}^n c_k u^k ~~~~~~{\rm with}\# \\
c_k=\frac{-(k+2)(k+1)}{2(n-k)} c_{k+2}-\frac{i \nu k (k+1)}{2 (n-k)}
c_{k+1} \# \\
c_{n-1}=\frac{-i \nu n(n-1) c_n}{2};~~c_n=2^n. 
\eea

To obtain our second solution to Eq.~(\ref{ldeq}), we make the change of
variables $z_l=-y_l/(\pi n)$, so that Eq.~(\ref{ldeq}) 
is transformed to:
\bea
\frac{n/d+y_l}{y_l}=2 \sum_{m \neq l} \frac{1}{y_l-y_m}
\eea
This is in a form where we can once again apply Stieltjes result. The
corresponding differential equation in this case is the associated
Laguerre equation 
Eq.~(\ref{Alag}). Thus $\{z_l\}$ are obtained as zeros of the associated
Laguerre polynomials $L^{(-n/d-1)}_{n}(-n \pi z)$ \cite{mutt}. Note
that the usual orthogonal 
Laguerre polynomials $L^a_n(y)$ have $ a > -1$, while in our case 
$ a=-(n/d+1) < -(n+1)$. In this case it can be proved \cite{szego}
that for $n$ even there are no 
real zeros while for odd $n$ there is a single real zero. Physically
this corresponds to the fact that the single-particle potential [see
App.~(\ref{appa})] in this
case is no longer confining and therefore we cannot get any
position of equilibrium for the particles.  

The two solutions described are related as:
\bea
Q_n(x)=\frac{(-i)^n n! 2^{n/2}}{(n/d)^{n/2}} L^{-n/d-1}_n( -i
(2n/d)^{1/2} x-n/d). 
\eea
We note that since in the limit $d \to 0$ we get $Q_n(x) \to H_n(x)$,
this leads to the following interesting identity relating 
the Hermite and Laguerre polynomials:
\bea
\lim_{d \to 0} \frac{(-i)^n n! 2^{n/2}}{(n/d)^{n/2}} L^{-n/d-1}_n(
-i(2n/d)^{1/2} x-n/d) = H_n(x).
\eea
Such a relation is of course well known, e.g. \cite{grad} for the case of large 
positive order $m$ in $L^{m}_n$

\section{Asymptotic distribution of zeros}
In the previous section it was shown that the solutions to
Eq.~(\ref{ldeq}) are the zeros of generalized associated Laguerre
polynomials. Unlike in the case of usual orthogonal polynomials where
the zeros are always real, in the present case the zeros are complex.
Szego \cite{szego} already notes that the zeros of $L_n^{-|m|}(z)$ are
 in general non-real except for at most one root. We found numerically that 
all the complex roots live on  a smooth curve in the complex plane, 
we now proceed to obtain the distribution of the roots in the
complex plane in the asymptotic limit $n \to \infty$.

Let us define the following Green's function (also known as the
Stieltjes transform in the mathematics literature):
\bea
G(z)=\frac{1}{n} \sum_{l=1}^n \frac{1}{z-z_l}.
\eea
An important aspect of our analysis is in the way in which we scale the
various arguments by $n$, and normalize by $n$ as above. The remaining 
variables, such as $G$, $z$ and $y$ below are of $O(1)$ as
 $n \rightarrow \infty$, whereby various simplifications arise for large $n$.

Using the results described in App.~(\ref{appa}) we get:
\bea
G(z)= \frac{1}{n F(z)} \frac{d F(z)}{d z}=\frac{1}{n} \frac{d}{dz}
\ln{(F(z))},
\label{dFdz}
\eea
where, in the present case, we have $F(z) = L_n^{-n/d-1}(-n \pi z)$.
The usual integral equation representation of the associated
Laguerre polynomial gives
\bea
F( z) &=& \frac{1}{2 \pi i} \oint \frac{\exp{(\frac{n \pi z
y}{1-y})}}{(1-y)^{-n/d} y^{n+1}} dy = \frac{1}{2 \pi i} \oint \frac{e^{-n
\phi(y,z)}}{y} dy,~~~~  \rm{where} \# \\
~\phi(y,z) &=& \ln{(y)}-\frac{1}{d} \ln{(1-y)}-\frac{\pi z y}{1-y}. 
\eea
The contour can be taken as any closed loop around the origin which
does not cross the branch line which we take as the real line from
$y=1$ to $\infty$. In the large $n$ limit the integral can be
evaluated by a saddle point method, as is standard in statistical physics.
The saddle point of  $\phi$ is determined for each $z$
We can choose the contour to pass through the appropriate
saddle point. In the usual case, this deformation leads to the
asymptotically exact  result for the ``free energy'', here we show that it
leads to the exact Green's function. The saddle-points are determined through 
$\p \phi(y,z)/ {\p y}= 0$ which gives
\bea
&& y_{\pm}=\frac{1-2d-\pi d z \pm \pi d
\sqrt{(z-z_{+})(z-z_{-})}}{2(1-d)}~~~~\rm{where}  \\
&& z_{\pm}=\frac{1-2 d }{\pi d} \pm \frac{i 2}{\pi} \sqrt{\frac{1}{d}-1}. \#
\eea
Out of the two saddlepoints we choose the one which gives a smaller value
for  $Re[\phi(y,z)]$ since that gives the dominant contribution to the
integral. Further it needs to be ensured that it is possible to
actually draw a contour, enclosing the origin and not crossing the
branch-cut,  such that, along the contour,  $\phi(y,z)$ takes its
minimum value at the saddlepoint. 
We find that either of the branches can be chosen depending on
the location of $z$ in the complex plane. The condition
\bea
Re[\delta \phi] = Re[\phi(y_{-},z)-\phi(y_{+},z)]=0 
\label{curve}
\eea 
determines the curve in $\angle z$ plane on which the
roots lie. This condition  actually determines two curves $C_1$ and
$C_2$, where $C_1$ concaves towards  the negative $Re[z]$ axis ( like
$x=-y^2$ ) and $C_2$ concaves in the other direction. The
curves $C_1$ and $C_2$ together  
form a closed region $R$ in $\angle z$ plane surrounding the point
 $(1-2 d)/( \pi d)$ [see Fig.~(\ref{halfd})]. To the right of the
curve $C_1$, the saddlepoint 
$y_+$ dominates. As we cross into the region $R$, the saddlepoint
$y_-$ takes over. However when we again cross the curve $C_2$, the
saddlepoint $y_-$ continues to determine the free energy even though
$\phi(y_+,z)$  is smaller. This is because it is no longer possible to
construct the necessary contour through $y_+$ [this is illustrated in 
Fig.~(\ref{contour})] and we remain stuck to $y_-$. 
Thus it is the curve $C_1$ which determines the locus of roots of the
polynomials. 

A brief discussion of the curves is in order. In terms of
 $\kappa= \sqrt{\frac{d}{1-d}}$ we replace $z$ by a  variable $\psi$ defined 
through $ z = \frac{1}{ \pi \kappa^2} [ ( 1- \kappa^2) + 2 \kappa \sinh( \psi) ]$
and hence the branch line of the square roots is a curve joining
the two points $\psi = \pm i \pi/2$. The difference in ``free energies''
 can be written as 
\bea
 \delta \phi = i \pi + 2 \psi + \frac{2}{\kappa} \cosh(\psi)
- \frac{ 1 + \kappa^2}{ \kappa^2} \log\{ \frac{ 1 + \kappa \exp (\psi)}
{1- \kappa \exp (-\psi)} \}. \label{bssnew1}
\eea
A particularly nice case is that of $d=1/2$ where we have $\kappa = 1$ and
hence another transformation to $b= (b_{re}+ i b_{im})$ defined through
 $\sinh(b) \sinh(\psi) = -1$ gives
\bea
\delta \phi= i \pi  + 2 [ b - \coth(b)].
\eea
The roots thus live on the curve $b_{re}= Re\coth(b_{re}+ i b_{im})$,
 which simplifies to the curve
\bea
b_{im} = \frac{1}{2} \arccos[ \cosh( 2 b_{re})- \sinh( 2
 b_{re})/b_{re} ]. \label{bssnew2} 
\eea
Note that $z= -2/(\pi \sinh(b))$, and $b_{re}$ varies between $bl=.00079$
 and $bu= 1.19968$ the values at which the argument of the $arccos$ hits 
$\pm1$. By choosing the appropriate branch of the $\arccos $ function we
obtain a smooth curve as depicted in Fig.~(\ref{halfd}).

In Fig.~(\ref{roots}) we show curves at different densities,
obtained by solving Eq.~(\ref{curve}) numerically. We also plot the
roots as obtained from an exact numerical solution of Eq.~(\ref{ldeq})
for $n=54$.

The asymptotic Green's function follows from Eq.~(\ref{dFdz}): 
\bea
G(z)=-{\left. \frac{d \phi(y,z)}
{d z} \right|}_{y^{\pm}}=\frac{\pi y_{\pm}}{1-y_{\pm}},
\eea 
where either of the branches $y_{\pm}$ is chosen depending on the
value of $z$. A second method of deriving the Green's function is
described in App.~(\ref{appb}).    

{\bf{Density of zeros}}: The Green's function contains complete
information on the roots and we now 
proceed to extract the density of zeros from it. We note that for the case
when the roots are all located on the real axis, the density of zeros
is given by the discontinuity in the imaginary part of the Green's
function across the branch line formed by the roots. Here we develop
a generalization of this procedure for the case when the roots are
distributed on a curve.

We first assume that the curve can be parametrized by a continuous
variable $s$ so that points on it are given by
$z(s)=(x(s),y(s))$. The greens function is given by
\bea
G(z)=\frac{1}{n} \sum_l \frac{1}{z-z(s_l)} \#.
\eea
We now define the following limiting function:
\bea
G_L(s)=\lim_{\n \to 0} G(z(s)+ \n)= \lim_{ \n \to 0} \byn
\sum_l \frac{1}{ z(s)-z(s_l)+\n};~~~~\n=\n_x+i \n_y  
\label{dfGl}
\eea 
Expanding around $s_l$ we get:
\bea
\lim_{\n \to 0} \frac{1}{
z(s)-z(s_l)+\n}&=& \lim_{\n \to 0}
\frac{1}{z'(s_l)}\frac{1}{[(s-s_l)+\hat{\n}]}~~~{\rm{where}} 
~~~~\hn=\frac{\n}{z'(s_l)} \# \\
&=& P- i \pi \frac{\d(s-s_l)}{z'(s_l)} sign(\hn_y),
\eea
where $P$ denotes the principal part having a continuous variation across the curve. 
Putting this into Eq.~(\ref{dfGl}) and using the definition
$\r(s)=\byn \sum \d(s-s_l)$ we get
\bea
G_L(s) &=& P-i \pi \frac{\rho(s)}{z'(s)}sign(\hn_y) \# \\
& = & P+ \frac{i \pi \rho(s)}{x'(s)+i y'(s)} sign[\n_x y'(s)-\n_y x'(s)].
\eea
We note that the above result has a simple geometric
meaning: the discontinuity is related to the variation
along the normal to the curve.
The discontinuity in the Green's function across the curve is thus
given by
\bea
G^{+}_L(s)-G^{-}_L(s)=\pm \frac{2 \pi i \r(s)}{z'(s)}=2 \pi (\a+ i\b),
\eea
where $\a$ and $\b$ define the real and imaginary parts of the jump.
Defining the projected densities $\r_x(x)=\byn
\sum \d(x-x_l)=\r(s)/{| x'(s) |}$ and $\r_y(y)=\byn \sum \d(y-y_l)=\r(s)/{|y'(s)|} 
$ we then get:
\bea
\r_x(x)=\frac{\a^2+\b^2}{|\b|};~~~~~\r_y(y)=\frac{\a^2+\b^2}{|\a|}.
\label{curvdns}
\eea
In Fig.~(\ref{dens}) we plot the density of zeros, $\r(s)$, obtained
numerically from 
Eqns.~(\ref{curve}) and (\ref{curvdns}), with the parameter $s$ chosen as
the Euclidean length along the curve. 

\section{Conclusions}
In summary we have studied a set of coupled nonlinear algebraic
equations which are essentially the Bethe ansatz equations for the
lowest energy states of the Heisenberg ferromagnetic chain, at small
densities. Both a perturbative and an exact solution of the equations
were obtained. In the former case we find that due to some very special 
properties of the Calogero matrix, we are able to calculate the
perturbation series to all orders. The exact solution of the equations
were obtained following a method due to Stieltjes and these are  
given by the zeros of generalized associated Laguerre polynomials. 
These zeros are distributed on the complex plane and using Green's
functions and saddle-point techniques we have obtained
the exact asymptotic distribution.
It may be noted that there are other examples in physics where one
needs to study complex zeros of polynomials as for example in the case
of the  Yang-Lee zeros. However this is probably one of the few examples where
the distribution of zeros in the complex plane has been computed exactly.

\appendix 
\section{}
\label{appa}
We give an account of Stieltjes electrostatic interpretation of zeros
of the orthogonal polynomials \cite{stiel,szego}. The case of the
Hermite polynomials is 
widely known in the physics literature and has been widely
applied. This does not seem to be the case for the other orthogonal
polynomials.
 
We first outline the derivation of the electrostatic interpretation for
the Hermite polynomial. Consider a set of charges in one-dimensions
interacting by a logarithmic repulsive force and confined within a
harmonic potential well. The potential energy of the system is given
by: 
\bea
E=\frac{1}{2} \sum_{l=1,n} x_l^2 -\frac{1}{2} \sum_{l \neq m}
\ln{\mid{x_l-x_m}\mid}
\label{eneq}
\eea
The minimum of the energy is given by
\bea
\p{E}/\p{x_l}=x_l-\sum_{m \ne l} \frac{1}{x_l-x_m}=0.
\label{mini}
\eea
Stieltjes result is that the $n$ zeros of the Hermite polynomial
$H_n(x)$ satisfy this equation. To show this, consider the $n$th-degree
polynomial 
\bea
g(x)=\prod_{l=1,n} (x-x_l),
\eea
where $x_l$ are the $n$ roots of Eq.~(\ref{mini}).
Taking derivatives with respect to $x$ we get
\bea
g'(x)=\sum_l \prod_{m \ne l} (x-x_m);~~~g''(x)=\sum_{k \ne l}
\prod_{m\ne k,l} (x-x_m).
\eea
Hence
\bea
\frac{g''(x_l)}{g'(x_l)}=2 \sum_{m \ne l} \frac{1}{x_l-x_m}=2 x_l
\label{zbyz}
\eea
since $x_l$ satisfy Eq.~(\ref{mini}). Thus we have $g''(x_l)-2 x_l
g'(x_l)=0 $. We further note that $f(x)=g''(x)-2 x g'(x)$ is a polynomial
of degree $n$ with the same roots as $g(x)$ and so they must be
identical apart from constant multiplicative factor. This constant
factor is determined by evaluating the coefficient of $x^n$ in
$f(x)$ which is $-2 n$. We thus obtain $f(x)=-2 n g(x)$ or
\bea
g''(x)-2 x g'(x)+2 n g(x)=0
\eea 
which is just the equation for the Hermite polynomials.
Thus it has been shown that the energy function in Eq.~(\ref{eneq}) is
minimized by a configuration in which the $n$ particles are located at
the zeros of the $H_n(x)$.

This result can be generalized by replacing $2 x_l$ in Eq.~(\ref{zbyz}) by
the expression $-q(x_l)/p(x_l)$ where $q(x)$ and $p(x)$ are arbitrary
polynomials of degree $1$ and $2$ respectively. Taking $q(x)=
q_0+q_1 x$ and $p(x)=p_0+p_1 x+p_2 x^2$ we repeat the previous
arguments to obtain
\bea
p(x) g''(x)+q(x) g'(x)+[-n(n-1) p_2- n q_1] g(x)=0.
\label{geneq}
\eea 
Denoting the polynomial solutions of this general equation by $g_n(x)$
we have the following result:
The zeros of the polynomial $g_n(x)$ correspond to the solution of
the set of equations 
\bea 
\frac{q(x_l)}{2 p(x_l)}+ \sum_{m \ne l} \frac{1}{x_l-x_m}=0.
\label{genst}
\eea
This corresponds to a minimization of the following $n$-particle
potential energy:
\bea
E=\frac{1}{2} \sum_l \int \frac{q(x_l)}{p(x_l)}+\frac{1}{2} \sum_{l \neq m} \ln{\mid x_l-x_m \mid}. 
\eea
We note that these results are valid for arbitrary complex values of
the coefficients $p_i$ and $q_i$, though a physical interpretation
cannot be given in all cases. In fact, as we shall now see, all the cases which
give rise to real confining potentials are precisely the ones
corresponding to the classical orthogonal polynomials. Physically we
expect that for a confining potential there will be real solutions to the
minimization problem and this is reflected in the fact that all
orthogonal polynomials of the $n$th degree have exactly  $n$ real
zeros (this of-course also follows from the Sturm-Liouville theory). We
now list the various cases which result in the standard orthogonal
polynomials. 	   

(i) $p_2 \ne 0$: In this case, it can be shown that upto linear and
scale transformations of the variable $x$, the most general form of
the differential equation which leads to a confining potential is  the
following:
\bea
(1-x^2) g''(x)+[b-a-(a+b+2)x] g'(x)+n(n+a +b+1) g(x)=0; ~~a,b>-1
\eea
The resulting potential $V(x)=-(1+a) \ln{(1-x)} - (1+b) \ln{(1+x)}$
confines particles within the domain $-1 < x <1$. This is the 
equation  for Jacobi polynomials, $P^{(a,b)}_n(x)$, of which the
Legendre and Chebyshev polynomials are special cases \cite{grad}.

(ii)$p_2=0,~p_1\ne 0$: In this case the most general form with a
bounding potential is
\bea
x g''(x) + (a+1-x) g'(x)+n g(x)=0;~~a > -1
\label{Alag}
\eea
The resulting potential $V(x)= x- (a+1) \ln{x}$ confines particles  to
$(0,\infty)$ and corresponds to the associated Laguerre polynomials,
$L^a_n(x)$ \cite{grad}.
 
(iii) Finally we have the case $p_1=p_2=0,~p_0 \ne 0$ and this gives
rise to Hermite's equation which has the harmonic potential
$V(x)=x^2/2$.

We note that apart from these special cases the other cases, where
$p_i, q_i$ are allowed to take arbitrary complex values, also may give
rise to important physical applications as indeed is so for the
example of Bethe roots considered in this paper. 

\section{}
\label{appb}
Here we derive a differential equation satisfied by the Green's
function. We also show how this also can be applied to the case of
the other orthogonal polynomials inorder to extract the density
of zeros in those cases.
We start with Eq.~(\ref{dFdz}) and take a single derivative to get:
\bea
n \frac{d G(z)}{dz}= -\frac{1}{F^2} (\frac{dF}{dz})^2+\frac{1}{F}
\frac{d^2F}{dz^2}   
\label{dGdz}
\eea
Now $F(z)=L_n^{-n/d-1}(-n \pi z)$ and so satisfies the following
second order equation (which is Eq.~(\ref{Alag}) with the substitution
$y=-n \pi z$ and $a+1=-n/d$):
\bea
z \frac{d^2F}{dz^2}+(-n/d+\pi n z) \frac{dF}{dz}-\pi n^2 F=0 \#
\eea
Using this and Eq.~(\ref{dFdz}) we can eliminate the
derivatives on the right hand side of Eq.~(\ref{dGdz}) to get our
equation for the Green's function:
\bea
\byn \frac{d G(z)}{dz}=-G^2(z)-(\frac{-1}{dz}+\pi) G(z) + \frac{\pi}{z}.
\eea
In the limit $n \to \infty$, the left hand side vanishes and the roots
of the quadratic on the right hand give us the two branches of the
Green's function. To choose between the branches ofcourse requires a
examination of the saddle points as described in section (IV).

This procedure of obtaining the asymptotic Green's function is easily
generalizable to other polynomial equations and we briefly discuss
a few applications to cases where the zeros are real.

(1) Associated Laguerre polynomials with $a+1=p n >0$: The
    replacement $d=-1/p$ at once gives us:
\bea
G(z)=\frac{-\pi z-p \pm \pi \sqrt{(z-x_{+}) (z-x_{-})}}{2 z} \# \\ 
\rm{with}~~~x_{\pm} =(-p-2\pm 2 \sqrt{p+1})/\pi.  
\eea
The roots are located in the region $ x_{-} < x < x_{+}$ and the
density of zeros, given by the discontinuity in the imaginary part of 
the Green's function is obtained as:
\bea
\rho(x)= \frac{1}{2 x} \sqrt{(x_{+}-x) (x-x_{-})} 
\eea

We can also consider the case where the coefficient $a$ does not scale
with $n$ but is a finite constant. The Green's function and density of
zeros are found to be independent of $a$:
\bea
G(z) &=& \frac{-\pi}{2}[ 1 - \sqrt{1+\frac{4}{\pi z}}] \# \\
\rho(x) &=& \frac{1}{2} \sqrt{\frac{4}{\pi |x| }-1}; ~~~-4/\pi < x <0  
\label{lag0}
\eea 

(2) Hermite polynomials: In this case we use scaled variables such
    that the zeros $z_l$ satisfy $H_n(\sqrt{n}z_l)=0 $. Proceeding as
    before we get the following equation for the Green's function:
\bea
\byn \frac{d G(z)}{dz}=-G^2(z)+2zG(z)-2. \#
\eea  
The asymptotic Green's function and the density of zeros follow
immediately:
\bea
G(z) &=& z-\sqrt{z^2-2} \# \\
\rho(x) &=& \frac{1}{\pi} \sqrt{2-x^2} 
\label{herm0}
\eea

(3) Jacobi polynomials: If $z_l$ satisfy $P^{(a,b)}_n(z_l)=0$, then
    the corresponding equation for the Green's function is
\bea
\byn \frac{d G(z)}{dz} = -G^2(z)-\frac{b-a-(a+b+2) z}{n(1-z^2)}
    G(z)-\frac{n+a+b+1}{n(1-z^2)}.  \# 
\eea
We thus get the following  asymptotic Green's function and density of
    states.
\bea
G(z) &=& \frac{1}{\sqrt{z^2-1}} \# \\
\rho(x) &=& \frac{1}{\pi \sqrt{1-x^2}}; ~~~~~-1 < x < 1 
\label{jaco0}
\eea
The results in Eq.~(\ref{lag0}), (\ref{herm0}) and (\ref{jaco0}) are
identical to those given in Refn. \cite{lang} where they have been
obtained by different methods.

\vbox{
\vspace{0.5cm}
\epsfxsize=8.0cm
\epsfysize=8.0cm
\epsffile{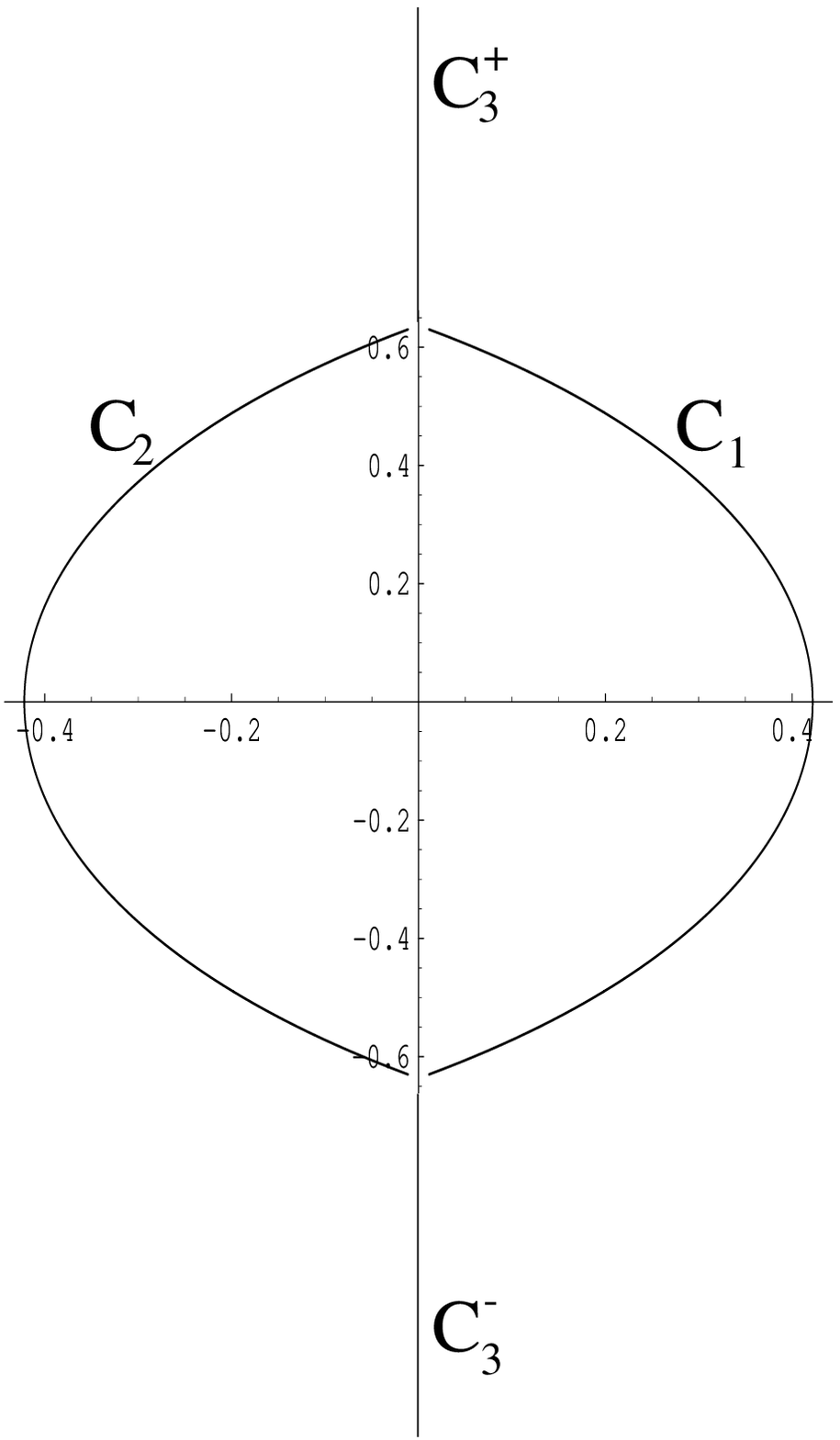}
\begin{figure}
\caption{\label{halfd} The curves $C_1$ and $C_2$, in the complex $z$
plane, for density $d=1/2$. At this density the curves are explictly
given by the formula in Eq.~(\ref{bssnew2}).  The curve $C_1$
corresponds to the true line of singularities on which the roots lie.
There is also a change of branch from $y^{+}$ to $y^{-}$ as we cross
the vertical lines $C_3^{\pm}$ from right to left. But there is no 
discontinuity of the Green's function across these lines and hence there
are no zeros on these lines.
}  
\end{figure}}
\vbox{
\vspace{0.5cm}
\epsfxsize=10.0cm
\epsfysize=8.0cm
\epsffile{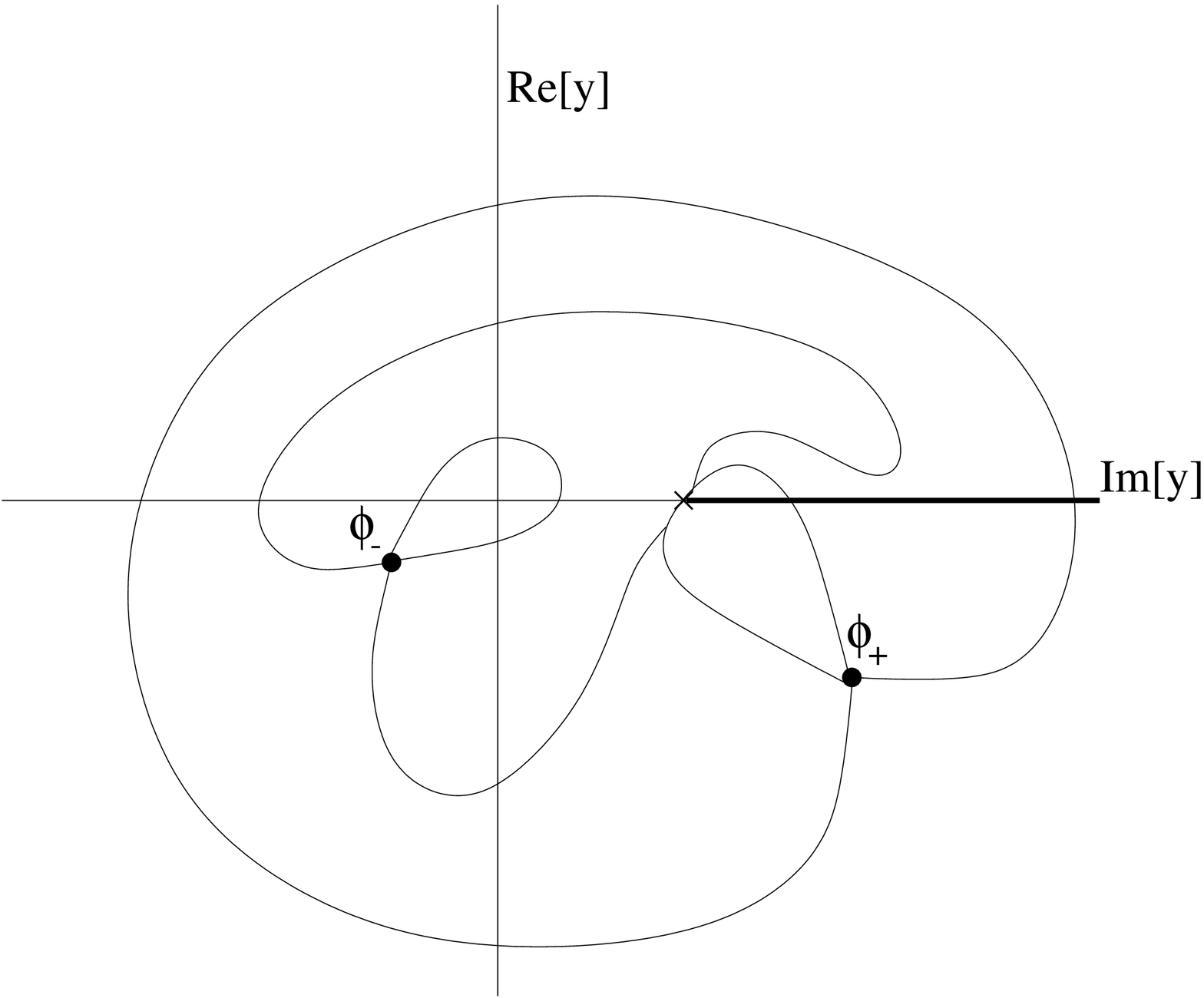}
\begin{figure}
\caption{\label{contour} The figure shows, schematically, contours of constant
$\phi_+$ and $\phi_-$ in the complex $y$ plane (with $d=0.6$ and
$z=-0.4+0.3 i$). For this case $\phi_+ < \phi_-$. It is clear that one
cannot draw the appropriate contour through the point $y_+$ and so we
are forced to choose the point $y_-$ even though it has a higher value
of $\phi$. This illustrates why there is no change in branch when we
cross the curve $C_2$.
}  
\end{figure}}
\vbox{
\vspace{0.5cm}
\epsfxsize=10.0cm
\epsfysize=8.0cm
\epsffile{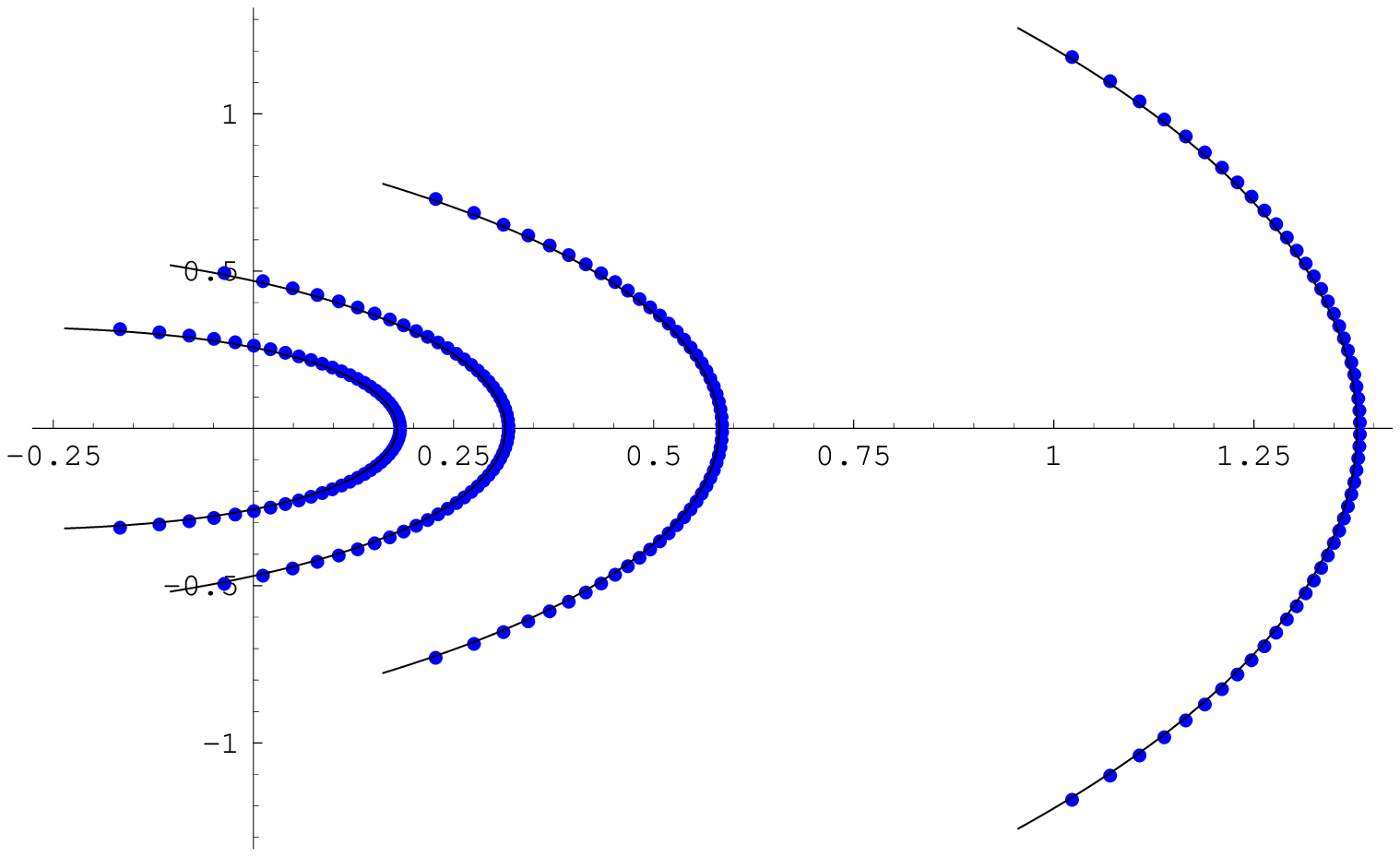}
\begin{figure}
\caption{\label{roots} The solid curves give the asymptotic
distribution of zeros in the 
complex $z$-plane obtained from a numerical solution of
Eq.~(\ref{curve}) for $d=0.2,0.4,0.6$ and $0.8$ ( extreme left ).
Also shown are numerically obtained roots (points) for $n=54$.
}  
\end{figure}}
\vbox{
\vspace{0.5cm}
\epsfxsize=10.0cm
\epsfysize=8.0cm
\epsffile{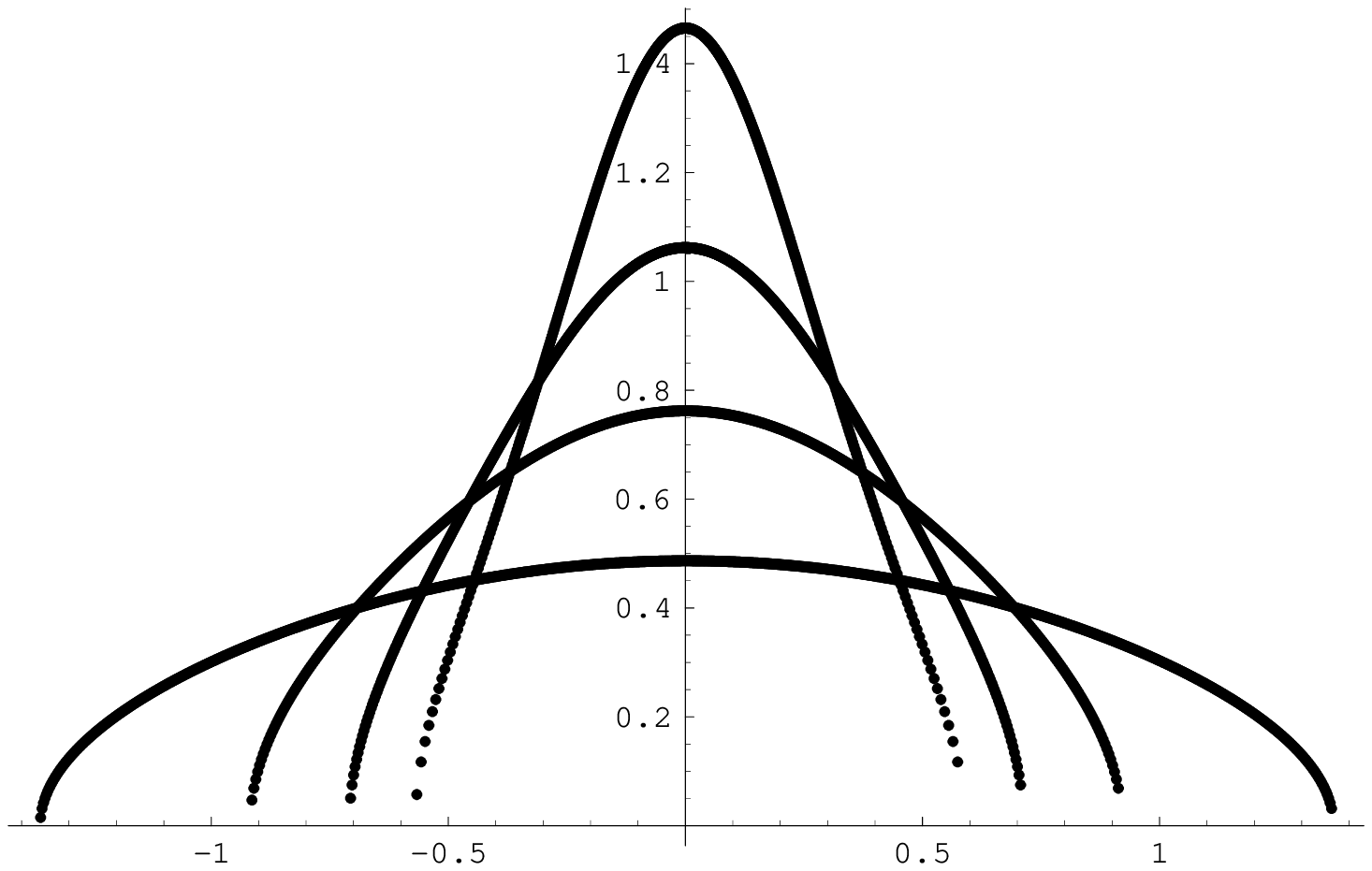}
\begin{figure}
\caption{\label{dens} 
The asymptotic density of zeros $\r(s)$ as a function of the Euclidean
length along the 
curve for $d=0.2,0.4,0.6$ and $0.8$ . The densities are evaluated
numerically using the formula in Eq.(~\ref{curvdns}).  
}  
\end{figure}}

\end{document}